\documentclass[times,review]{elsarticle}

\usepackage{framed,multirow}
\usepackage{amsmath}

\newcommand{\mywidth}{0.5\textwidth}

 \usepackage{amsmath}
\usepackage{siunitx}
 \usepackage{graphicx}
\usepackage{subfigure}

 \usepackage{float}

 \journal{Journal of Computational Physics}

\begin{document}

\begin{frontmatter}

\title{Convolutional discrete Fourier transform method for calculating thermal neutron cross section in liquids }%

\author[1,2]{Rong Du}
\author[1,2]{Xiao-Xiao Cai\corref{cor1}}
\ead{caixx@ihep.ac.cn}
\cortext[cor1]{Corresponding author:}

\address[1]{Institute of High Energy Physics, Chinese Academy of Sciences}
\address[2]{Spallation Neutron Source Science Center}

\begin{abstract}
  Being exact at both short- and long-time limits, the Gaussian approximation is widely used to calculate neutron incoherent inelastic scattering functions in liquids.
However, to overcome a few numerical difficulties, extra physical approximations are often employed to ease the evaluation.

  In this work, a new numerical method, called convolutional discrete Fourier transform, is proposed to perform Fourier transform of $\exp[-f(t)]$.
We have applied this method to compute the differential cross sections of light water up to \SI{10}{\eV}.
The obtained results, thoroughly benchmarked against experimental data, showed a much higher dynamic range than conventional fast Fourier transform.
The calculated integral cross sections agree closely with the light water data in the state-of-the-art nuclear data library.
It is in evidence that this numerical method can be used in the place of the extra physical approximations.
\end{abstract}

\end{frontmatter}

\section{Introduction}

The advances of molecular dynamics  simulations~\cite{Rahman1971,Toukan1985,Marti1996}  provided valuable insights into the  microscopic structural and dynamic properties of liquids. It has been shown that for the application of thermal neutron scattering, accurate neutron incoherent inelastic scattering in water can be calculated using molecular dynamics simulated density of states without any free parameter~\cite{Noguere2021}. Such  progress implies that it is now possible to predict the incoherent process reliably prior to any experiment.

However, there exist numerical challenges when evaluating the cross sections numerically using the Gaussian approximation~\cite{Rahman1962}, for instance, the highly oscillatory and singular integrand.
To tackle the problem, physical approximations are often employed to ease such a computation.
Under these approximations, motions of atoms are strictly categorised as a particular type of diffusion and solid-like vibrations.
Therefore, available numerical routines, such as the phonon expansion method~\cite{Sjolander1958}, which are valid for systems that move around equilibrium positions, can be directly used.  
The final numerical results of the combined models are often consistent with experimental cross sections in many liquids~\cite{Noguere2021}.

Apart from that, the main difficulty of the computation lies within the finite numerical dynamic range of Fourier transforms. Discrete Fast Fourier transform is typically used to correlate atomic microscopic structure and dynamics with measurable scattering cross sections, i.e. a function of momentum and energy transfers that can span tens of orders of magnitudes.
However, this numerical procedure is known to have spectral leakage~\cite{Harris1978} from a frequency bin to the others close by. 
Therefore, a weak and fast decaying signal can be submerged by the leakage from an intense bin, resulting in a significantly smaller dynamic range than the precision that digitised floating-point number can offer. Even when the evaluation of the cross section is assisted with physical approximations, at large energy transfer, e.g. above \SI{1}{\eV}, the numerical error can no longer be ignored. A short-collision-time (STC) approximation is often employed to replace the noise-contaminated results~\cite{MacFarlane2010}.

%Typically, a large numerical dynamic range is required to represent a scattering function, and the requirement is especially demanding for liquids.
%Satisfying the principle of detailed balance, scattering functions contain a large distortion factor $\exp(-\hbar\omega/k_bT)$, where $\hbar\omega$ is the energy transfer between neutron and the target system, and $T$ is the temperature of the system.
%This factor is one of the main responsible for the fast decaying behaviour of a scattering function when energy transfer becomes large.
%Moreover, diffusive motions in liquids contribute a shape peak to the scattering function allocated around zero transfer. At small momentum transfer, this so-called quasi-elastic peak spans several orders of magnitude.

%For neutron slowing down simulations,  the  function is inadequate, even if the best window function is used for minimising the spectral leakage~\cite{a paper about window}.

In this work, we attempt to solve the numerical difficulties using mathematical techniques solely, a convolutional  discrete Fourier transform (CDFT) method is proposed as an alternative to the three physical approximations introduced earlier.

In section~\ref{sLiquid}, the theory of neutron scattering and Gaussian approximation, along with the physical approximations for easing numerical calculation, are introduced.
The convolutional discrete Fourier transform (CDFT) method is firstly introduced in section~\ref{sHDRFT},
and applied to the Gaussian approximation in section~\ref{ssLiquidHDRFT}.
Using the identical input as the state-of-the-art CAB light water cross section~\cite{Damian2013, Damianpriv}, the CDFT calculated scattering function of light water is benchmarked against differential and integral cross sections in section~\ref{sBenchmark}. 
This work is concluded in section~\ref{sConclusion}.

\section{Neutron incoherent Inelastic scattering in liquids}
\label{sLiquid}

\subsection{Gaussian approximation}

In Van Hove's space-time correlations~\cite{VanHove1954}, the scattering function $S(Q,\omega)$ is given by the Fourier transform of the intermediate scattering function $F(Q,t)$, which can be approximated by a Gaussian~\cite{Rahman1962}
\begin{align}
	S(Q,\omega) &= \frac{1}{2\pi} \int e^{-i \omega t} F(Q,t) \mathrm{d} t \nonumber\\
	&= \frac{1}{2\pi} \int e^{-i \omega t}  \exp\left[- \frac {Q^2}{2}\Gamma(t)\right]  \mathrm{d} t \label{fGaussian}
\end{align}
%$F(Q,t)$ is defined as the Fourier transformation of $G(r,t)$ in space $\vec{r}$.
%\begin{equation}
%	F(Q,t)=\int e^{i\vec{Q} \cdot \vec{r}} G(r,t) \mathrm d\vec{r}
%\end{equation}

%\begin{equation}\label{sqw_dir}
%	S(Q,\omega) = \frac{1}{2\pi} \int e^{-i \omega t}
%	\exp\left(- \frac {Q^2}{2}\Gamma(t)\right) \mathrm{d} t
%\end{equation}

With the fluctuation-dissipation theorem, it has been shown that, $\Gamma(t)$  in liquids can be expressed as ~\cite{Rahman1962}
\begin{equation}\label{gamma}
	\Gamma(t) = \frac{\hbar}{m} \int_0^{\infty} \mathrm{d}\omega \frac{\rho(\omega)}{\omega}
	\left[\coth\left(\frac{\hbar \omega}{2k_B T}\right)
	(1-\cos {\omega t})-i \sin{\omega t}\right]
\end{equation}
Here, $\rho(\omega)$ is the density of states, $k_B$ is the Boltzmann constant.

\subsection{Physical models assisted numerical evaluation}
\label{ssSolid}

Physical models are often employed to facilitate the evaluation of Eq.~\ref{fGaussian}.
This section introduces the conventional formulation for the evaluation briefly,  a more detailed discussion can be found in~\cite{CAB2014}.
The key technique for such evaluation is to break down the cumbersome width function $\Gamma(t)$ into two easier-to-evaluate components, i.e. a diffusion part and a solid-like vibration part.

\begin{equation}\label{vdos}
		\rho(\omega) = w_{\mathrm{d}}\rho_{\mathrm{diff}}(\omega) + w_{\mathrm{v}}\rho_{\mathrm{vib}}(\omega)
\end{equation}
Where $w_{\mathrm{d}}$ and $w_{\mathrm{v}}$ are the weights satisfy $w_{\mathrm{d}}+w_{\mathrm{v}}=1$.
The diffusion component is simply described by the Egelstaff-Schofield model~\cite{Egelstaff1962}
\begin{equation}
	\begin{split}
		\rho_{\mathrm{diff}}(\epsilon) &= \frac{4cw_t}{\pi \hbar kT} \sqrt{c^2+1/4}
		 \sinh{\frac{\epsilon}{2kT}}
		  K_1\left( \frac{\epsilon}{kT} \sqrt{c^2+1/4}\right)
	\end{split}
\end{equation}
where $c$ and $w_t$ are the diffusion constant and translational weight, respectively.
\begin{equation}
	c=\frac{M_{\mathrm{diff}}D}{\hbar}
\end{equation}

The diffusion coefficient $D$ that equals $\pi k T \rho(0)/2m$ is correlated linearly with $\rho(0)$, and $M_{\mathrm{diff}}$ is the diffusion mass.

For the solid-like part, the model that describes neutron-phonon interaction in harmonic crystals are used~\cite{Sjolander1958}.

\begin{align}\label{solidF}
	F(Q,t) &= \exp{\left[-\frac{Q^2}{2} \gamma(0)\right]}\exp{\left[\frac{Q^2}{2} \gamma(t)\right]} \\
	&= \exp(-2W)\exp{\left[\frac{Q^2}{2} \gamma(t)\right]}
\end{align}
where $\exp(-2W)$ is the Debye-Waller factor, and
\begin{equation} \label{fsolidgamma}
	\gamma(t) = \frac{\hbar}{m} \int_0^{\infty} \mathrm{d}\omega \frac{\rho_{\mathrm{vib}}(\omega)}{\omega}
	\left[\coth\left(\frac{\hbar \omega}{2kT}\right)\cos{\omega t}+i\sin{\omega t} \right]
\end{equation}

Towards zero energy, for any three-dimensional solid material, the density of states always decay according to a power law~\cite{Sears1995}.
Therefore, the integrand of Eq.~\ref{fsolidgamma} has no singularity at $\omega = 0$.
Indicating by Eq.~\ref{vdos}, to make this combination of physic approximations work, it is crucial to subtract a suitable portion of the diffusive density of states, i.e. $\rho_{\mathrm{diff}}$, from the original liquid state to obtain an $\rho_{\mathrm{vib}}$ that satisfies the power law at small energies, i.e. $a\omega^2$ .

At large energy and momentum transfers, the scattering function became a Gaussian without any detailed structures. Due to the numerical difficulties, conventional evaluation  is likely to fail in that region, hence the short-collision time approximation is often used.

\begin{equation} \label{eSCT}
	S(Q, \omega) = \sqrt{\frac{1}{4\pi k_b T E_r}} \exp\left[ -\frac{(\hbar\omega - E_r)^2}{4\pi E_r}  \right] \,\,\,\,\, \mathrm{, where} \,\,\,\,\, E_r=\frac{\hbar^2 Q^2}{2m}
\end{equation}

\section{The convolutional discrete Fourier transform (CDFT) method}
\label{sHDRFT}

The concept of the CDFT method is based on the Taylor expansion of the function for Fourier transform.
We discuss the case that the function of interest is an exponential function $\exp[-f(t)]$, and $f(t)>0$.
The Fourier transform of that can be expanded as
\begin{align}\label{fft}
	F(\omega) &= \int e^{-i\omega t} \exp[-f(t)] \mathrm{d} t \\
	&= \sum_{n=0}^{\infty} \frac{1}{n!} \int e^{-i\omega t} [-f(t)]^n \mathrm{d} t \\
\end{align}

From the Convolution theorem, it is straight forward to have
\begin{equation}\label{ConvTheo}
\int e^{-i\omega t}  f_1(t) f_2(t) \mathrm{d} t =  F_1(\omega)  \otimes   F_2(\omega)
\end{equation}
Therefore, it is obvious to see that the higher order terms of the summation can be obtained by convoluting the results of lower order terms.
However, this method is not numerically stable. The integral area of the convoluted result is the produce of the areas of the two input functions. After a great number of convolutions, the results can exceed the limiting value that can be expressed by a floating number. Therefore, it is desired to use normalised functions as input.

According to Bochner's theorem, the Fourier transform of a probability measure on $f(t)$ is necessarily a normalized continuous positive-definite function.
For a finite discrete function, to make the function continuous in the overall domain of $t$, it is required to change the function to make it zero at both boundaries.
For an event function, it is obvious to get $r(t)=f_{max}-f(t)$. Eq.~\ref{fft} can then be rearranged as
\begin{equation}\label{eConvFirstStep}
	\begin{split}
		F(\omega) &= e^{-f_{max}} \int e^{-i\omega t} e^{r(t)} \mathrm{d} t \\
		&= e^{-f_{max}} \sum_{n=0}^{\infty} \frac{r^n(0)}{n!}  \int e^{-i\omega t} \left[ \frac{r(t)}{r(0)} \right]^n \mathrm{d} t 	\end{split}
\end{equation}

%We use $h(t)$ to instead of $\vert t \vert$ or $t^2$, and assume $r(t)=x-h(t)$. The Eq.~\ref{fft} becomes:
%\begin{equation}
%	\begin{split}
%		F(\omega) &= e^{-ax} \int e^{-i\omega t} e^{ar(t)} \mathrm{d} t \\
%		&= e^{-ax} \int e^{-i\omega t} \sum_0^{\infty} \frac{a^n}{n!} r^n(t) \mathrm{d} t \\
%		&= e^{-ax} \sum_0^{\infty} \frac{a^n r^n(0)}{n!} \int e^{-i\omega t}
%		\left[\frac{r(t)}{r(0)}\right]^n \mathrm{d} t
%	\end{split}
%\end{equation}

Denoting $g_n(\omega)$ function as
\begin{equation}\label{gnw}
 	g_n(\omega)=\int e^{-i\omega t}  \left[ \frac{r(t)}{r(0)} \right]^n  \mathrm{d} t
 \end{equation}

It can be see immediately that $g_0$ is a $\mathrm{sinc}$ function and not involved in the convolution procedure.
On the other hand, the integral of $g_1$ is normalised (see Eq.~\ref{g1normal}) and is the fundamental input of the convolutions.
 \begin{equation}\label{g1normal}
 \begin{split}
	\int g_1(\omega) d \omega &= \iint e^{-i \omega t} \frac{r(t)}{r(0)} \mathrm{d} t  \mathrm{d}\omega \\
	&= \int \delta (t) \frac{r(t)}{r(0)} \mathrm{d} t \\
	&= 1
 \end{split}
 \end{equation}

Therefore, the integral of higher order $g$ functions are also unity.

As $r(0)=f_{max}$, Eq.~\ref{eConvFirstStep} becomes
 \begin{equation}
	 F(\omega) = e^{-f_{max}} \sum_0^{\infty} \frac{(f_{max})^n}{n!} g_n(\omega)
	 %F(\omega) = e^{-ax} \sum_0^{\infty} \frac{x^n a^n}{n!} g_n(\omega)
\end{equation}

Until now, the numerical procedure is straight forward. Here we introduced the distortion factor, which is used to maximise the dynamic range of the convolution.

We defined an distort function as $e^{aw}$. It has the relation with the convolution as
\begin{equation}\label{eRelation}
	\begin{split}
	F_1(\omega) \otimes F_2(\omega) &= \int F_1(\mu) \cdot F_2(\omega-\mu) d\mu \\
		&= \int e^{a\mu} \hat{F_1}(\mu) \cdot e^{a(\omega-\mu)} \hat{F_2}(\omega-\mu) d\mu \\
		&= e^{a\omega} \int \hat{F_1}(\mu) \cdot \hat{F_2}(\omega-\mu) d\mu \\
		&= e^{a\omega}\hat{F_1}(\mu) \otimes \hat{F_2}(\omega-\mu)
	\end{split}
\end{equation}

Denoting
\begin{equation}
	\hat{g}_n(\omega)  = e^{-a\omega} g_n(\omega)
\end{equation}

Using Eq.~\ref{eRelation}, we finally have,
\begin{equation}\label{fft_final}
	F(\omega) = \exp(-f_{max}+a\omega) \sum_{n=0}^{\infty} \frac{(f_{max})^n}{n!} \hat{g}_n(\omega)
	%F(\omega) = \exp(-ax+y\omega) \sum_{n=0}^{\infty}\frac{x^n a^n}{n!}\hat{g}_n(\omega)
\end{equation}
where $a$ is a constant, which only affects the dynamic range of the final convolution. To optimise the result for a higher dynamic range, the CDFT method evaluates the Fourier transform by convoluting the distorted function $\hat{g}_n(\omega)$, instead of the original function $g_n(\omega)$.
The significant improvement of such procedure is introduced in section~\ref{sBenchmark}.

\subsection{Applying the CDFT Method to the cross section calculation}
\label{ssLiquidHDRFT}
	Recall that
\begin{equation}\label{esqwmaster}
	S(Q,\omega) = \frac{1}{2\pi} \int e^{-i \omega t}
	\exp\left(- \frac {Q^2}{2}\Gamma(t)\right) \mathrm{d} t
\end{equation}

Using Eq.~\ref{eConvFirstStep}, the integration becomes
\begin{equation}\label{eFFTequv}
		S(Q,\omega) = \exp\left({-\frac{\Gamma_{max}Q^2}{2}}\right)
		\sum_{n=0}^{\infty} \left(\frac{\Gamma_{max}Q^2}{2}\right)^n \frac{1}{n!}
		\int e^{-i\omega t}\left[\frac{r(t)}{r(0)}\right]^n \mathrm{d} t
\end{equation}
where $r(t)=\Gamma_{max}-\Gamma(t)$. Numerically, this equation is equivalent to the direct fast Fourier transform.

Based on Eq.~\ref{fft_final}, the integration can also be expressed as
\begin{equation}\label{eCDFT}
	S(Q, \omega) = \exp{\left(-\frac{\Gamma_{max}Q^2}{2}+a \omega\right)}
	\sum_{n=0}^{\infty} \left(\frac{\Gamma_{max}Q^2}{2}\right)^n \frac{1}{n!} \hat{g}_n(\omega)
\end{equation}
Note that for our typical calculations are room temperature for light water, we have found that the distortion coefficient $a=\omega/3k_b T$ works quite well.

%Fig.~\ref{fig:gnw} compares the 11th term of Eq.~\ref{eFFTequv} and Eq.~\ref{eCDFT} when $Q = \si{1}{\AA^{-1}}$. Noise floor can not be observed when distortion factor is applied.
%\begin{figure}[H]
%\centering
%	\includegraphics[width={\mywidth}]{distort.pdf}
%\caption{Evaluation results the 11th term of Eq.~\ref{eFFTequv} and Eq.~\ref{eCDFT} for light water scattering function calculation, $Q = \si{1}{\AA^{-1}}$.}
%\label{fig:gnw}
%\end{figure}

At large momentum transfer, the factor  $\exp (\Gamma_{max} Q^2)$ in Eq.~\ref{eCDFT} may overflow. To prevent that, we apply the convolutional method again.
Denoting $h(\omega)=S(Q,\omega)$, at a given $Q$ where the scattering function is obtained by Eq.~\ref{eCDFT}, according to Eq.~\ref{esqwmaster}, the scattering function at higher Q values can be calculated as 
\begin{equation}\label{finalsqw}
	 S(\sqrt{2^{n-1}} Q ,\omega) = \underbrace{h(\omega) \otimes h(\omega) \dots \otimes h(\omega)}_{n}
	 % S(Q,\omega) = \underbrace{a \ a}
\end{equation}

Notice that the self-scattering function satisfies the sum-rule $\int S(Q,\omega) \mathrm{d} \omega=1$, hence a large number of $h(\omega)$ functions can be convoluted without encountering the numerical overflowing problem.
The distortion factor, similar to Eq.~\ref{eCDFT}, are also applied in our numerical implementation.

\section{Results and discussion}
\label{sBenchmark}

\subsection{Evaluation of the self-scattering function}
\begin{figure}[H]
\centering
\includegraphics[width={\mywidth},page=1]{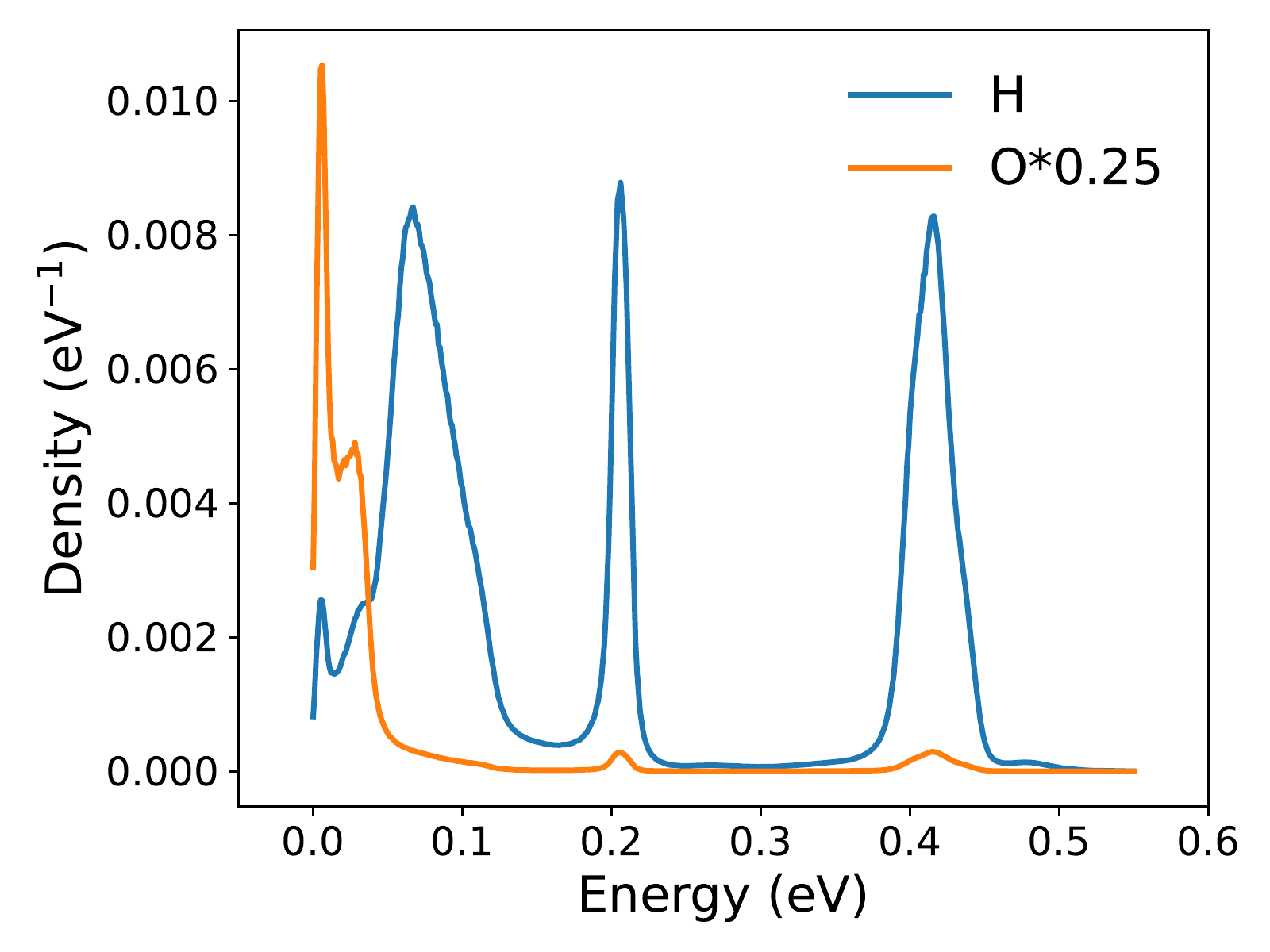}
\caption{The density of states for hydrogen and oxygen in light water for CAB model at \SI{297}{\K}~\cite{Marquez2013}. }
\label{fig:vdos}
\end{figure}

The density of states (DOS) for both hydrogen and oxygen in light water for the CAB model are used as the input of our calculation~\cite{Marquez2013}.
The data, shown in Fig.~\ref{fig:vdos}, are provided by the one of the authors of the CAB model~\cite{Damianpriv}.

\begin{figure}[H]
\centering
\includegraphics[width={\mywidth},page=2]{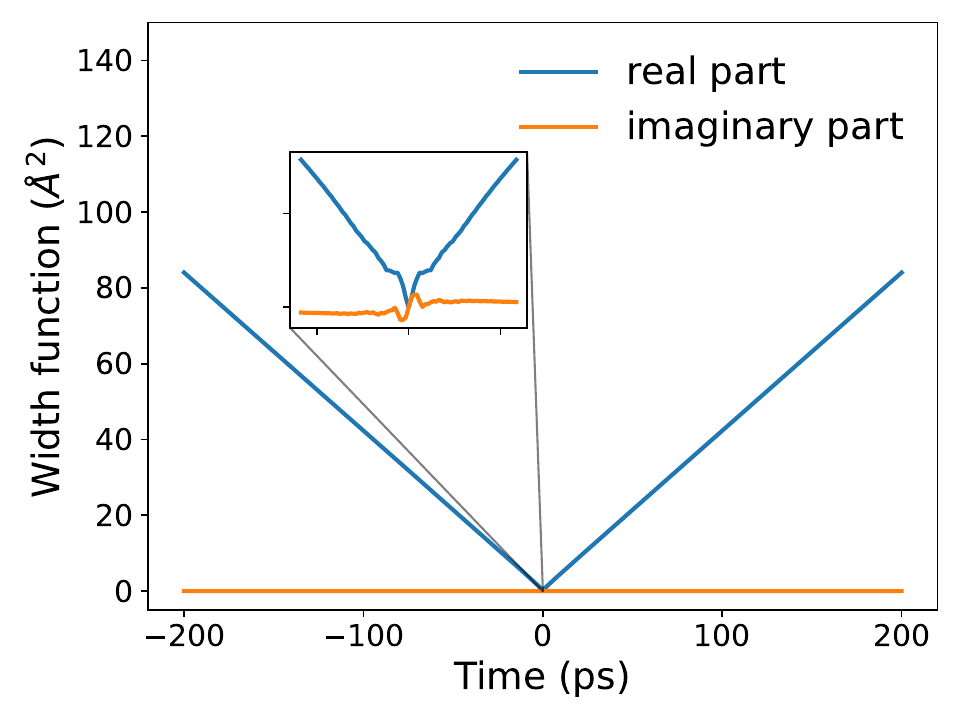}
\caption{Width function $\Gamma(t)$ for hydrogen at \SI{297}{\K}.}
\label{fig:gamma}
\end{figure}

%gamma
The $\Gamma(t)$ functions defined in Eq.~\ref{gamma} are evaluated using the Filon method~\cite{Filon1968} in the range between \si{-200}\si{\pico\second} and \si{200}\si{\pico\second},
of which the detailed implementation can be found in~\ref{sFilonMethod}. 
For light water, a molecular dynamic simulation of \si{100}\si{\pico\second} is able to capture the important slow dynamics for scattering function~\cite{CAB2014}, the time range chose in this work is considered to be adequate.
The result of $\Gamma(t)$ for hydrogen at \SI{297}{\K} is shown in Fig.~\ref{fig:gamma}.
Complex fast dynamics are  concentrated in the region less than \si{1}\si{\pico\second}.
The real part of $\Gamma(t)$ increases linearly when the physics is mainly governed by the diffusion process at large times.

\begin{figure}[H]
\centering
\includegraphics[width={\mywidth}]{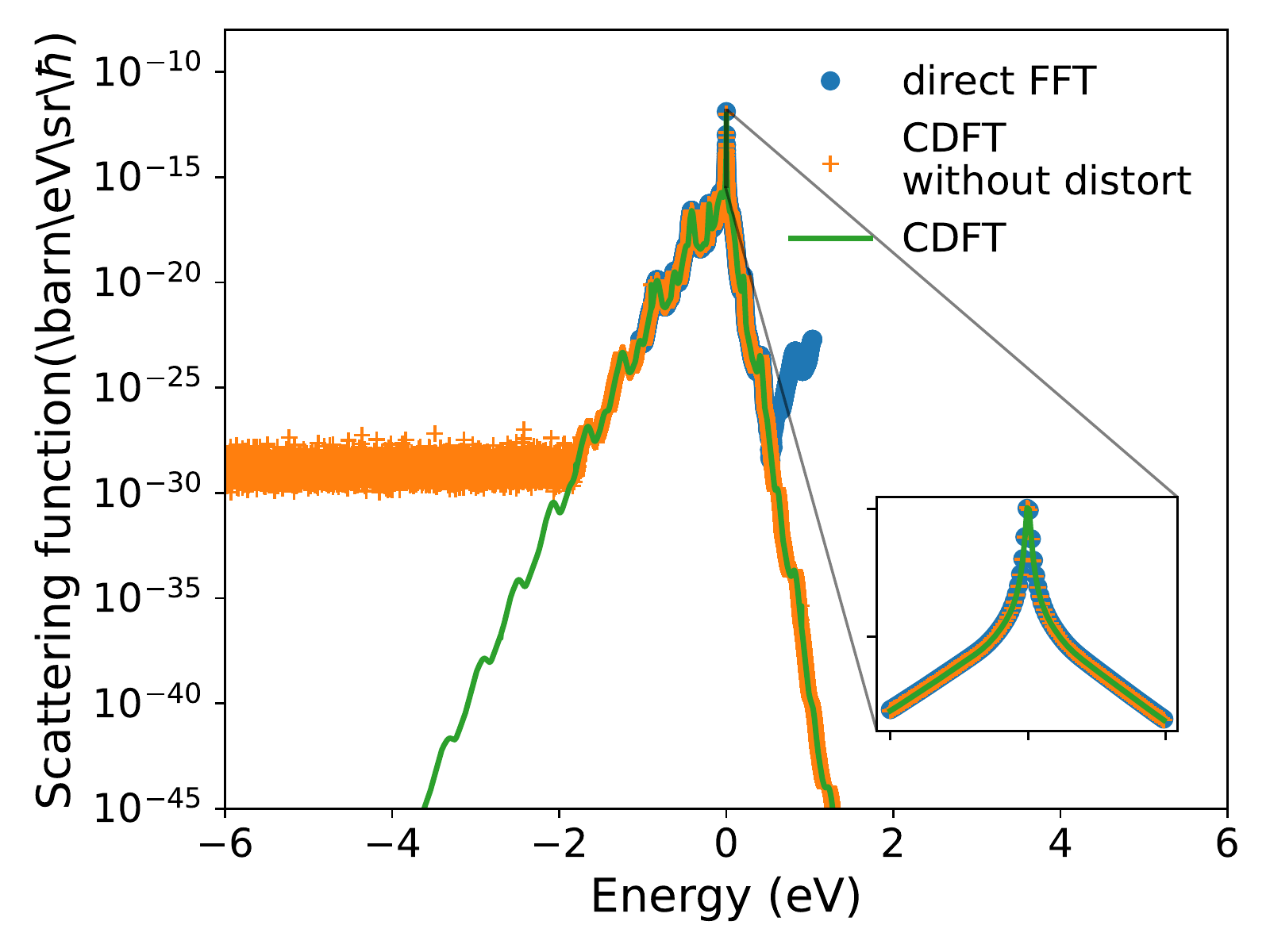}
\caption{Comparison of calculated scattering function from direct fast Fourier transform (i.e. Eq.~\ref{esqwmaster}), CDFT without distortion (i.e. Eq.~\ref{eFFTequv}) and with distortion (i.e. Eq.~\ref{eCDFT}), $Q=\si{1}\si{\AA^{-1}}$  }
\label{fig:sqw_smallQ}
\end{figure}

Fig.~\ref{fig:sqw_smallQ} compares the numerical results of three different methods for the Fourier transform. The reduced momentum transfer is \si{1}\si{\AA^{-1}}.
At small energy transfers, results from three methods are all agreed. When the absolute value of the momentum transfer is greater than \num{2}\si{\eV}, the scattering function calculated by the CDFT method showed a much higher dynamic range than the others.

\begin{figure}[H]
\centering
\begin{minipage}[t]{0.45\linewidth}
	\includegraphics[width=\textwidth, page=2]{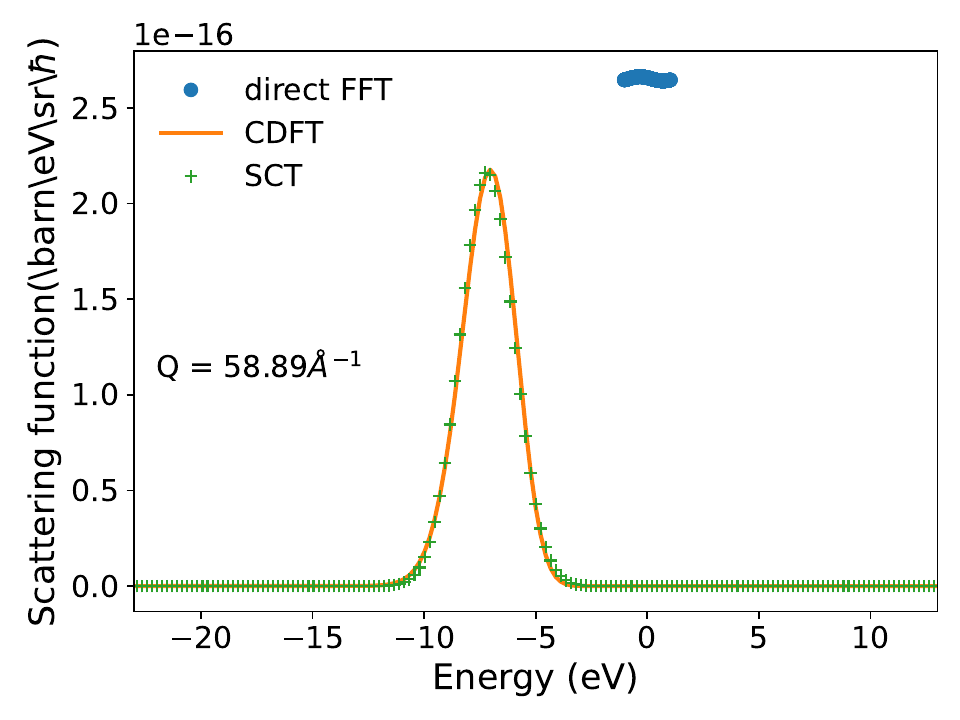}
\end{minipage}
\begin{minipage}[t]{0.45\linewidth}
	\includegraphics[width=\textwidth, page=3]{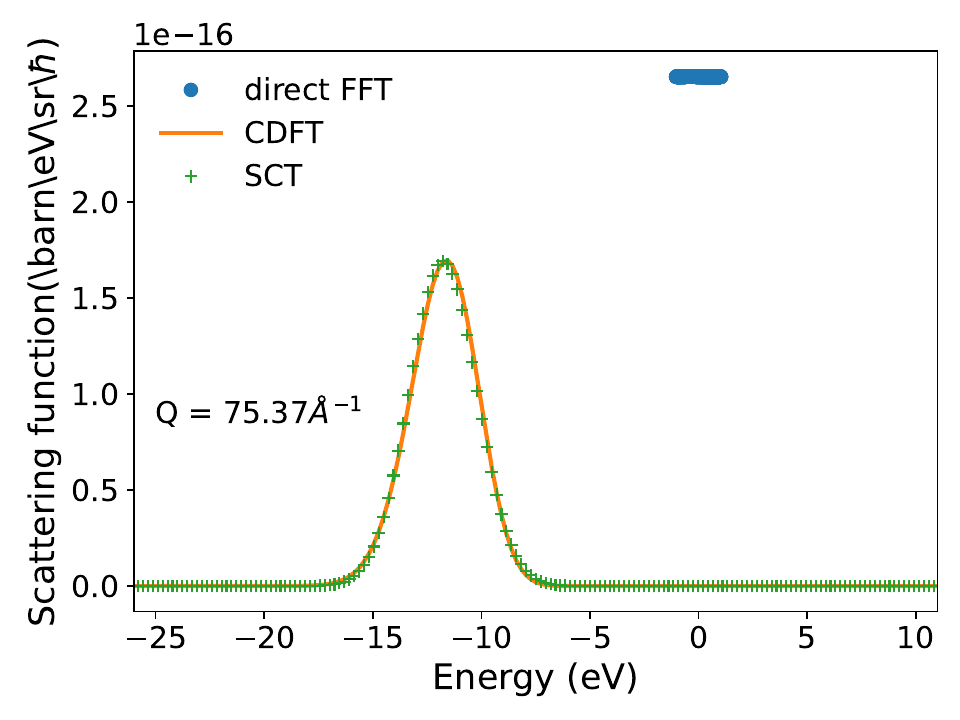}
\end{minipage}
\caption{Comparison of calculated scattering function from direct fast Fourier transform (i.e.~\ref{esqwmaster}), CDFT with distortion (i.e. Eq.~\ref{finalsqw}) and the prediction of the SCT approximation (i.e. Eq.~\ref{eSCT}), at larger Q.}
\label{fig:sqw_largeQ}
\end{figure}

When the momentum transfer is very large, the short-collision-time (SCT) approximation becomes valid.
Fig.~\ref{fig:sqw_largeQ} compares the numerical results from direct fast Fourier transform, CDFT method and the prediction of the SCT model, which is only valid when moment transfer is high.
The result from the CDFT method showed good agreement with the prediction of SCT method. On the other hand, the results from direct fast Fourier transform agree with neither of the other results, suggesting such a method could suffer from significant error when the momentum transfer is high.

\begin{figure}[H]
\centering
\includegraphics[width={\mywidth}]{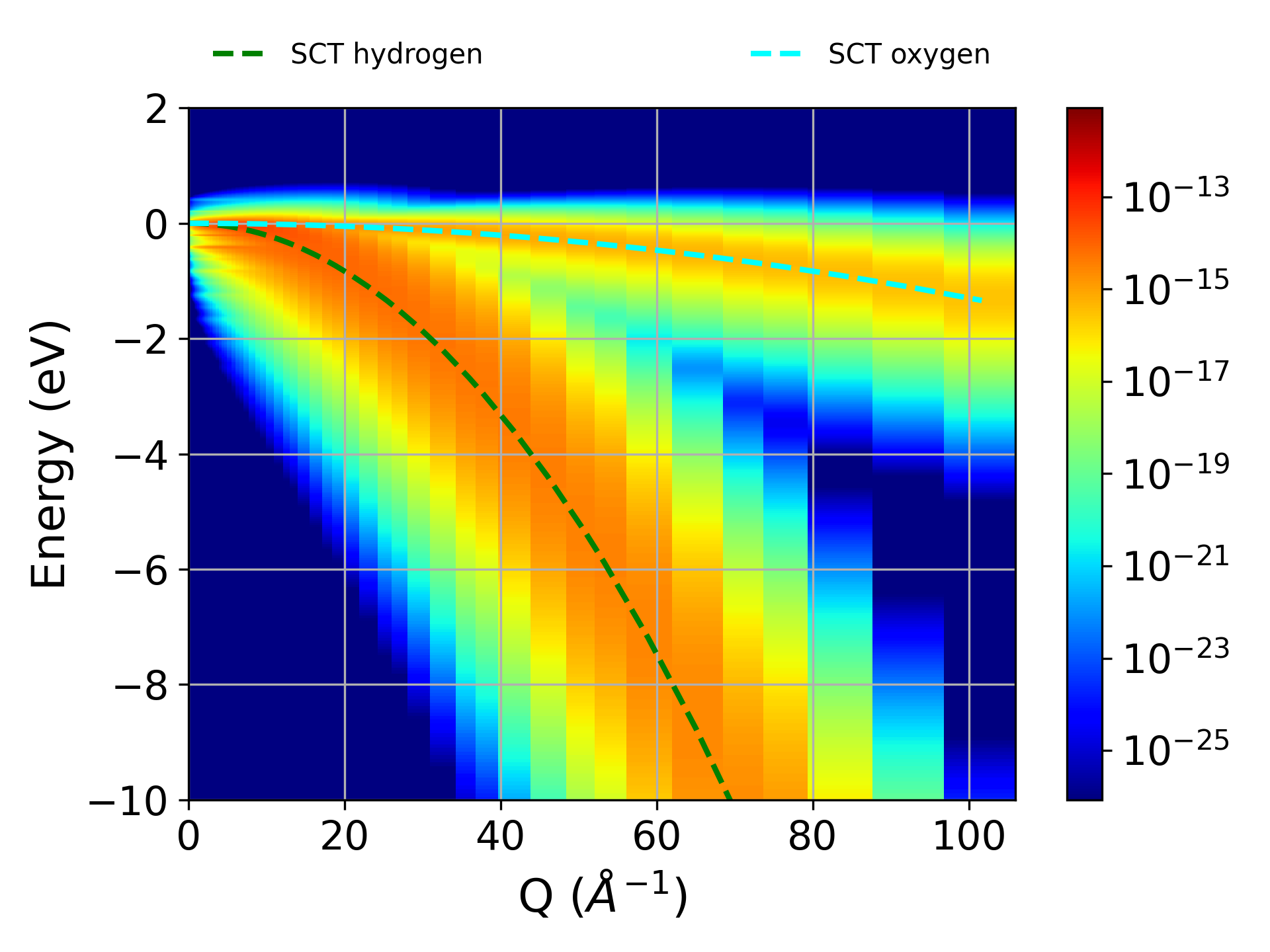}
\caption{The scattering function of light water at \SI{297}{\K} calculated using Eq.~\ref{eCDFT} and Eq.~\ref{finalsqw} }
\label{fig:hotSqw}
\end{figure}

The complete scattering functions for light water molecule is shown in Fig.~\ref{fig:hotSqw}. For the generation of the data, Eq.~\ref{eCDFT} and Eq.~\ref{finalsqw} are used to calculate the function below and above $Q=\SI{0.3}{\AA^{-1}}$, respectively.
The peak position of the SCT approximation (i.e. Eq.~\ref{eSCT}) at given Q are shown as broken lines. These lines are well aligned with the peaked intensities  contributed by hydrogen and oxygen.

\subsection{Quasi-elastic scattering}

\begin{figure}[H]
\centering
\includegraphics[width={\mywidth}]{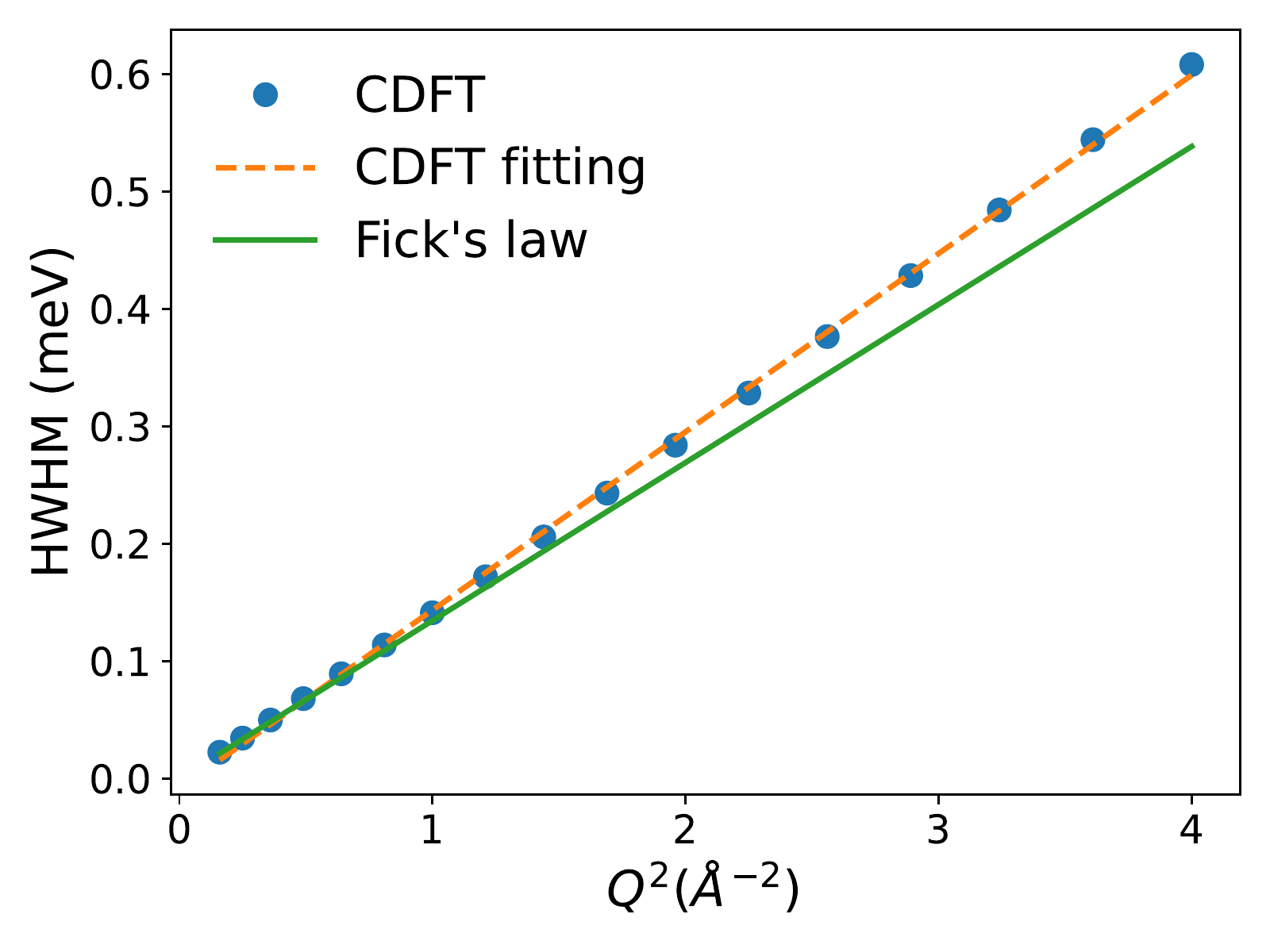}
\caption{Comparison of the HWHM from CDFT and Fick's law.}
\label{fig:hwhm}
\end{figure}

Fig.~\ref{fig:hwhm} compares the half width at half maximum (HWHM) of the scattering function at small momentum transfers. Our calculated results are slightly greater than that from a simple isotropic three-dimensional diffusion model (or the Fick's law, see for example chapter 5.4 in~\cite{Squires2012}).
According to that model, the scattering function at small energy and momentum transfers can be described as a Lorentzian
\begin{align}\label{Lorentzian}
S(Q,\omega) = \frac{1}{\pi \hbar}\frac{DQ^2}{(DQ^2)^2+\omega^2}
\end{align}

\begin{figure}[H]
\centering
\includegraphics[width={\mywidth}]{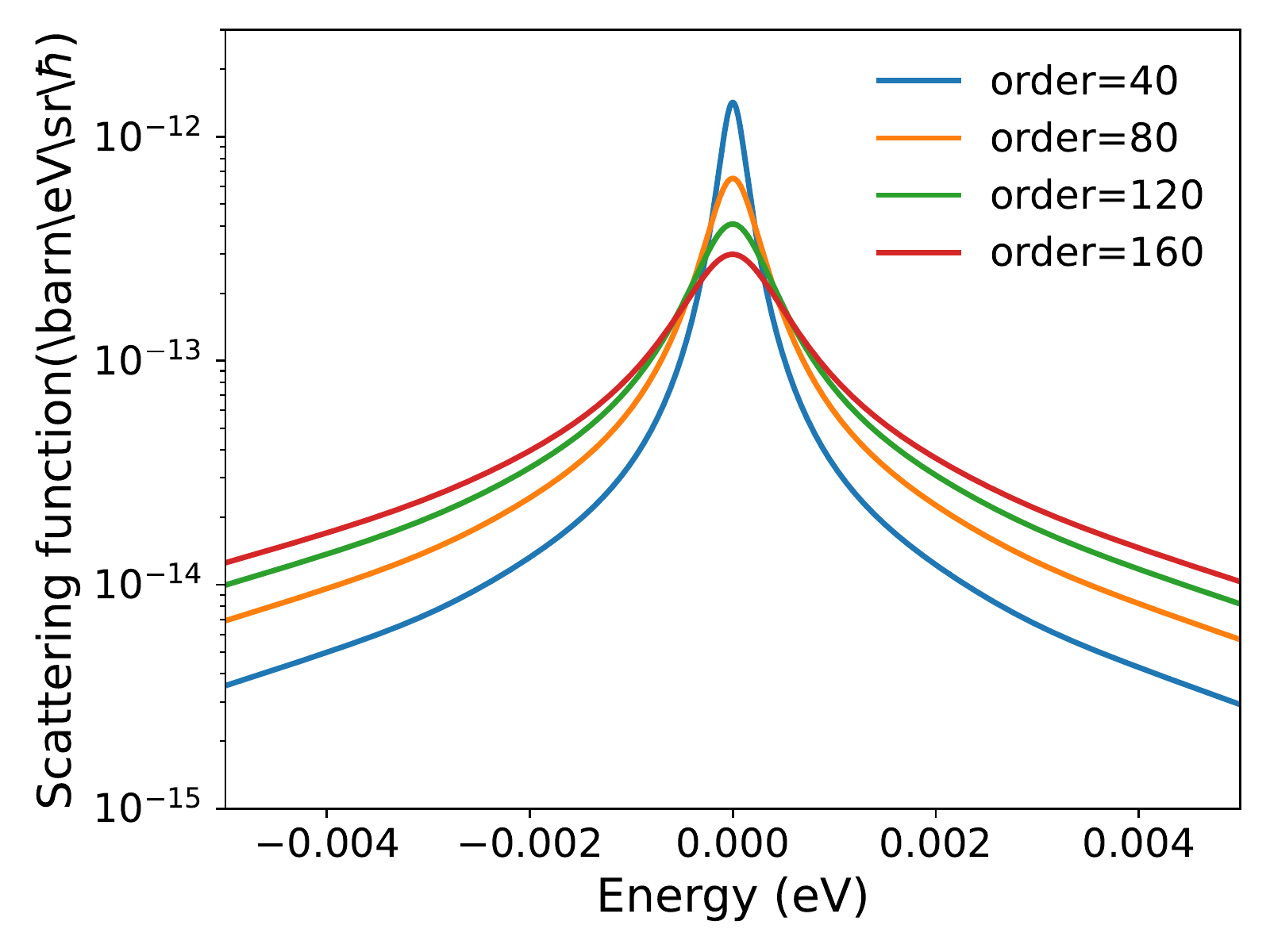}
\caption{Comparison of scattering function with different scattering orders, $Q=1$.}
\label{fig:sqwOrder}
\end{figure}

The discrepancies are because, unlike the Fick's law, the HWHM from our calculation is not contributed by the diffusive process alone~\cite{Levesque1970}. Fig.~\ref{fig:sqwOrder} shows the contributions to the scattering function from different scattering orders. At small energies, multiple scatterings of diffusive motion and energetic phonons can also contribute to the energy region that the diffusive peak allocated. For example, in a two phonon scattering case, a neutron absorbs and emits phonons of very similar energies. 
As can be observed in the figure, the contribution from higher order scattering is with a broader HWHM.

The intensity of those peaks grow quickly at greater momentum transfer and the shape diffusion peak will eventually be submerged by the multiple scattering contributions. In that case, the short-collision-time approximation becomes valid.

\subsection{Benchmarking against experimental data}
\begin{figure}[H]
\centering
\begin{minipage}[t]{0.45\linewidth}
	\includegraphics[width=\textwidth, page=3]{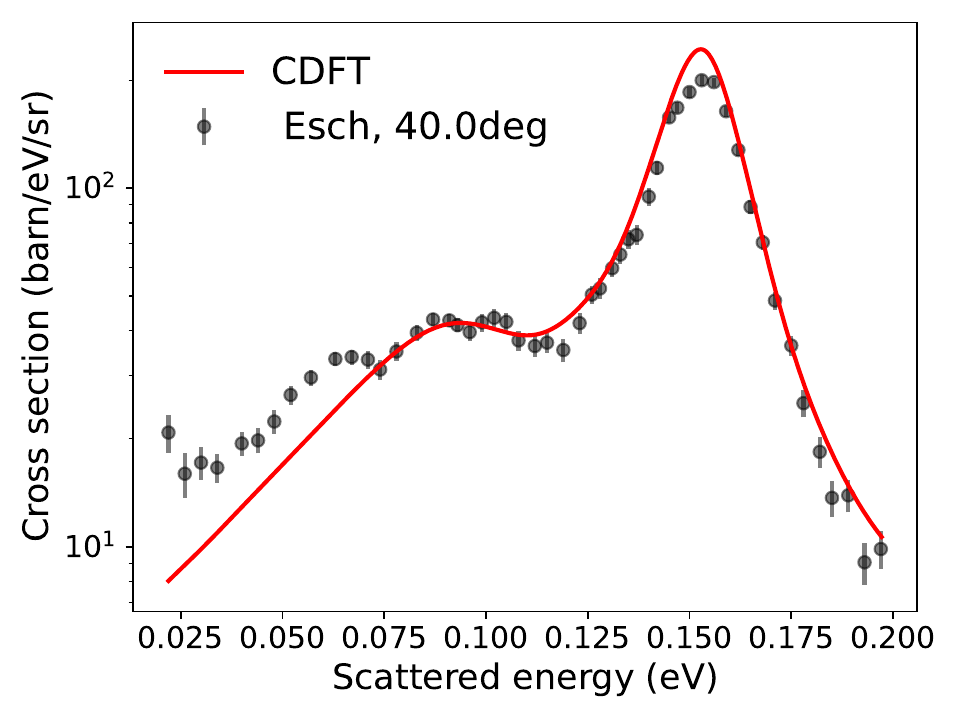}
\end{minipage}
\begin{minipage}[t]{0.45\linewidth}
	\includegraphics[width=\textwidth, page=4]{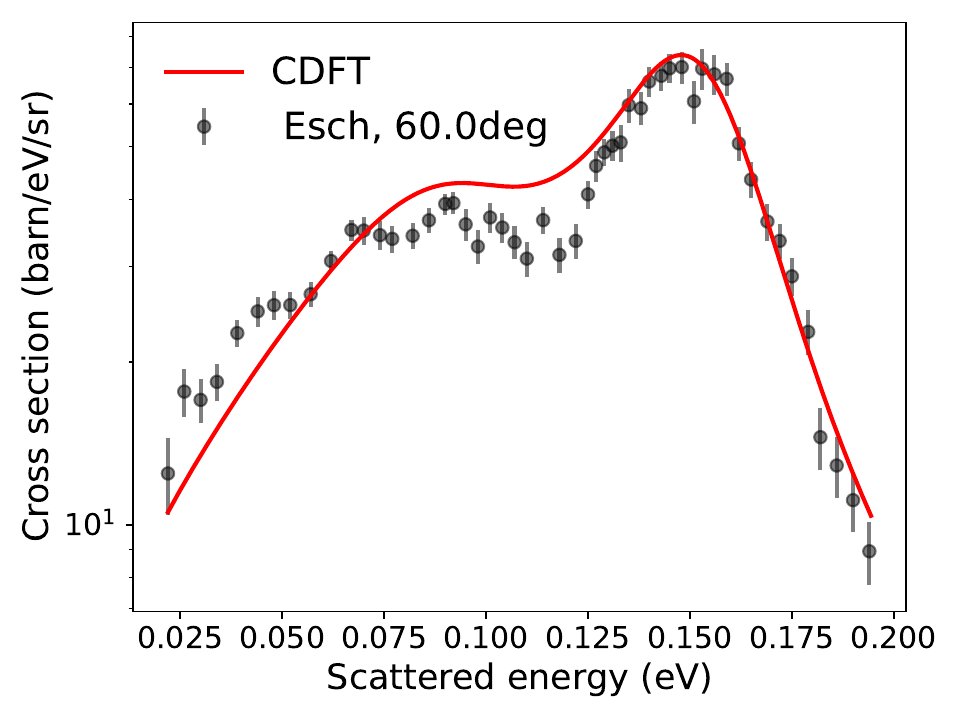}
\end{minipage}
\caption{Double differential scattering cross section for light water, $E_0 = \SI{0.154}{\eV}$.}
\label{fig:ddxs_154}
\end{figure}

\begin{figure}[H]
\centering
\begin{minipage}[t]{0.45\linewidth}
	\includegraphics[width=\textwidth, page=1]{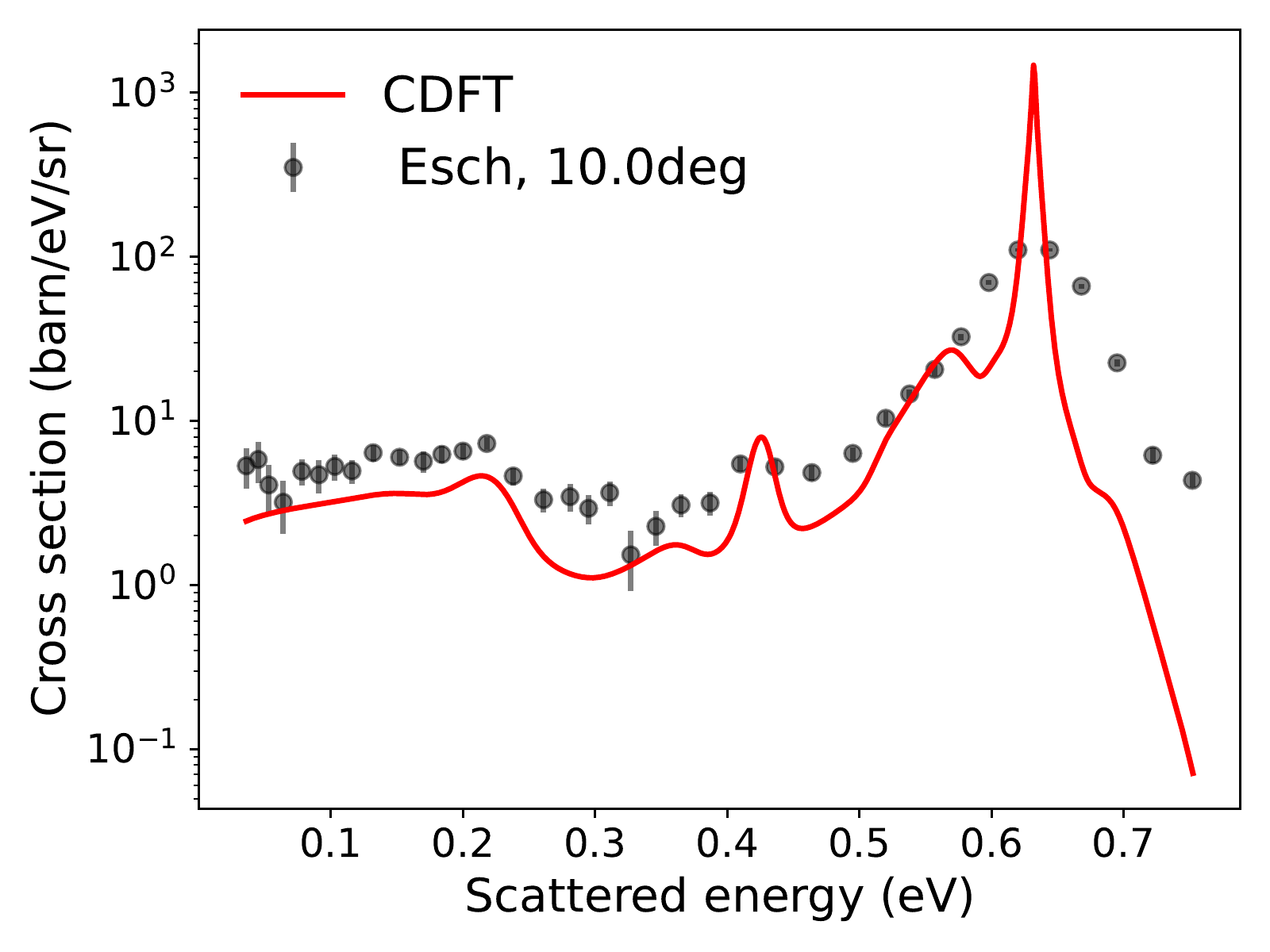}
\end{minipage}
\begin{minipage}[t]{0.45\linewidth}
	\includegraphics[width=\textwidth, page=2]{ddcs.pdf}
\end{minipage}
\caption{Double differential scattering cross section for light water, $E_0 = \SI{0.632}{\eV}$ .}
\label{fig:ddxs_632}
\end{figure}

The double differential scattering cross section for light water is computed and compared with experiment results from Esch~\cite{Esch1971}.
A comparison for $E_0 = \SI{0.154}{\eV}$ and $E_0 = \SI{0.632}{\eV}$  with different angles were shown in Fig.~\ref{fig:ddxs_154} and Fig.~\ref{fig:ddxs_632}. Excellent agreements are observed.

\begin{figure}[H]
\centering
\includegraphics[width={\mywidth}]{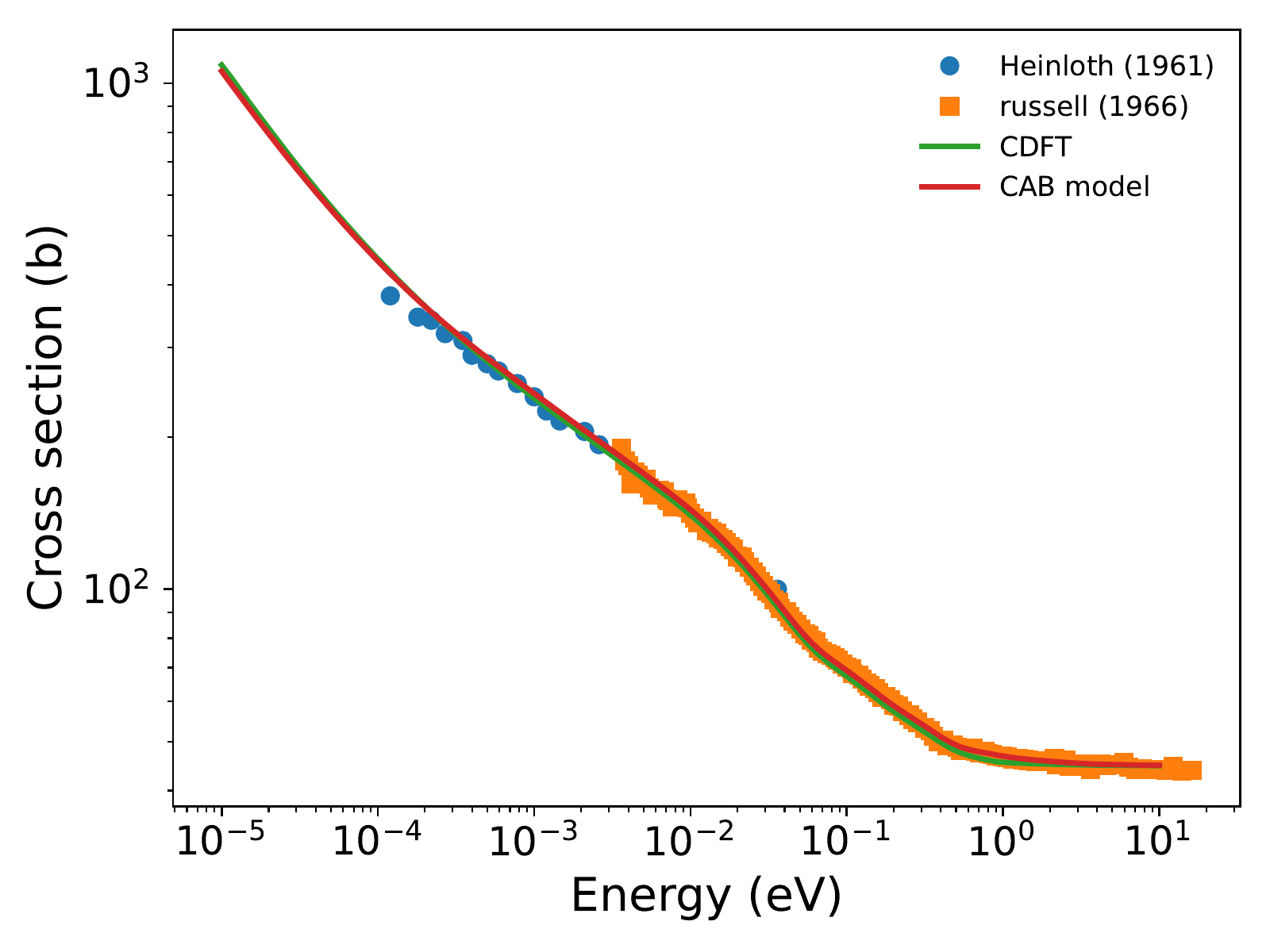}
\caption{Total cross section for light water at \SI{297}{\K}.}
\label{fig:totxs}
\end{figure}

	Total cross section is calculated from \SI{1}{\mu\eV} to \SI{10}{\eV} at \SI{297}{\K} to assess  the accuracy of the CDFT method.
Fig.~\ref{fig:totxs} shows the comparison of several results from both theoretical calculation and experiment, which are from Heinloth (1961)~\cite{Heinloth1961} and russell (1966)~\cite{Russell1966}.
Along with the CAB model cross section, the total cross section from the CDFT method are in good agreement with experimental data.

\section{Conclusion}
\label{sConclusion}
A numerical procedure, called convolutional discrete Fourier transform (CDFT) method, for evaluating the Fourier transform of an exponential function is proposed. We showed that this method can accurately evaluate neutron incoherent inelastic scattering function in light water. The computed scattering function is consistent with the state-of-the-art CAB model. Instead of physical approximations assisted evaluation of the scattering function, the CDFT method relies on no additional physical approximations. The results from the CDFT method should provide more confidence when working with a new liquid, on with the validity of the commonly used physical approximations are unknown.

The CDFT method will be released in an open source package, along side with the NCrystal project~\cite{Cai2020}.

\section*{Acknowledgements}
This research is supported by the National Natural Science Foundation of China (Grant No.12075266). The authors are grateful to Dr. Damian, J. I. Marquez for valuable discussions and substantial support.

\appendix
%\section{approve zero order}
%
%The scattering law for liquid is defined by
% \begin{equation}
% 	S(Q, \omega) = \frac{1}{2 \pi \hbar} \int [G(\vec{r}, t) - \rho]
% 	\exp(i\vec{Q} \cdot \vec{r} - i \omega t) d\vec{r} dt
% \end{equation}
% where $\rho=\frac{N}{V}$ is number density. After integral in time, scattering function is
%
% \begin{equation}
% 	\begin{split}
% 		S(\vec{Q}) &= 1+\int [g(\vec{r})-\rho] e^{i\vec{Q}\vec{r}}d\vec{r} \\
% 		&= 1+\frac{4 \pi}{Q} \int_0^{\infty}[g(\vec{r})-\rho] \sin{Qr}r dr
% 	\end{split}
% \end{equation}
% But for gaussian approximation, scattering function includes the delta function of elastic peak.
% \begin{equation}
% \begin{split}
%  I_{el}(\vec{Q},\omega) &= \int \rho e^{i \vec{Q} \cdot \vec{r}} d\vec{r} \\
%  &= \rho \frac{1}{(2\pi)^3}\delta(\vec{Q})
%  \end{split}
% \end{equation}
% And the elastic part of scattering function is
%
% \begin{equation}
% \begin{split}
% 	S_{el}(\vec{Q},\omega) &= \frac{1}{2\pi \hbar} \int \rho
% 	\frac{\delta{\vec{Q}}}{(2\pi)^3} e^{-i \omega t} dt \\
% 	&= \frac{\rho \delta{\vec{Q}} \delta{\omega} }{(2\pi)^4 \hbar}
% \end{split}
% \end{equation}
% The elastic peak contribute to $S(\vec{Q},\omega)$ when $Q$ and $\omega$ are both zero.
%

\section{Filon method}
\label{sFilonMethod}
The Filon method is used for the integrals of the form:
\begin{equation}
	\begin{split}
		S &= \int_a^b f(x) \sin(kx) dx \\
		C &= \int_a^b f(x) \cos(kx) dx
	\end{split}
\end{equation}
The equations for numerical implementation are these:
\begin{equation}
	S=h[\alpha(f_0\cos(kx_0)-f_{2p}\cos(kx_{2p}))+\beta S_{2p}+\gamma S_{2p-1}],
\end{equation}
\begin{equation}
	C=h[\alpha(f_{2p}\sin(kx_{2p})-f_0\sin(kx_0))+\beta C_{2p}+\gamma C_{2p-1}],
\end{equation}
where
\begin{equation}
	S_{2p}=\sum_{i=0}^{p} f_{2i}\sin(kx_{2i})-
		\frac{1}{2}[f_0\sin(kx_0)+f_{2p}\sin(kx_{2p})],
\end{equation}
\begin{equation}
	S_{2p-1}=\sum_{i=1}^{p} f_{2i-1}\sin(kx_{2i-1}),
\end{equation}
\begin{equation}
	C_{2p}=\sum_{i=0}^p f_{2i}\cos(kx_{2i})-
		\frac{1}{2}[f_0\cos(kx_0)+f_{2p}\cos(kx_{2p})],
\end{equation}
\begin{equation}
	C_{2p-1}=\sum_{i=1}^p f_{2i-1}\cos(kx_{2i-1}),
\end{equation}
\begin{equation}
	\alpha = 1/\theta + \sin(2\theta)/{2\theta^2}-(2\sin^2{\theta})/\theta^3,
\end{equation}
\begin{equation}
	\beta = 2[(1+\cos^2{\theta})/\theta^2-\sin(2\theta)/\theta^3],
\end{equation}
\begin{equation}
	\gamma = 4(\sin{\theta}/\theta^3-\cos \theta/\theta^2),
\end{equation}
\begin{equation}
	\theta = kh ,
\end{equation}
\begin{equation}
	f_i=f(x_i), x_{i+1}-x_i=h, x_0=a, x_{2p}=b
\end{equation}

\bibliographystyle{ieeetr}
%\bibliography{ref.bib}

 \end{document}